\documentclass[%
 reprint,
 superscriptaddress,
 amsmath,amssymb,
aps,
]{revtex4-1}

\usepackage[export]{adjustbox}

\usepackage{graphicx}
\usepackage{epstopdf}
\usepackage{dcolumn}
\usepackage{bm}
\usepackage{color}
\usepackage[caption=false]{subfig}
\usepackage{xspace}
\usepackage{epsfig}
\usepackage{multirow}
\usepackage[]{units}
\usepackage{textcomp}
\usepackage{hyperref}

\usepackage[displaymath]{lineno}

\usepackage{cleveref}
\crefname{figure}{fig.}{figs.}
\crefname{table}{tab.}{tabs.}
\crefname{equation}{eq.}{eqs.}

\input symbols

\begin{document}

\preprint{XXX}

\title{Search for short baseline $\nue$ disappearance with the T2K near detector}


\newcommand{\INSTC}{\affiliation{University of Alberta, Centre for Particle Physics, Department of Physics, Edmonton, Alberta, Canada}}
\newcommand{\INSTEE}{\affiliation{University of Bern, Albert Einstein Center for Fundamental Physics, Laboratory for High Energy Physics (LHEP), Bern, Switzerland}}
\newcommand{\INSTFE}{\affiliation{Boston University, Department of Physics, Boston, Massachusetts, U.S.A.}}
\newcommand{\INSTD}{\affiliation{University of British Columbia, Department of Physics and Astronomy, Vancouver, British Columbia, Canada}}
\newcommand{\INSTGA}{\affiliation{University of California, Irvine, Department of Physics and Astronomy, Irvine, California, U.S.A.}}
\newcommand{\INSTI}{\affiliation{IRFU, CEA Saclay, Gif-sur-Yvette, France}}
\newcommand{\INSTGB}{\affiliation{University of Colorado at Boulder, Department of Physics, Boulder, Colorado, U.S.A.}}
\newcommand{\INSTFG}{\affiliation{Colorado State University, Department of Physics, Fort Collins, Colorado, U.S.A.}}
\newcommand{\INSTFH}{\affiliation{Duke University, Department of Physics, Durham, North Carolina, U.S.A.}}
\newcommand{\INSTBA}{\affiliation{Ecole Polytechnique, IN2P3-CNRS, Laboratoire Leprince-Ringuet, Palaiseau, France }}
\newcommand{\INSTEF}{\affiliation{ETH Zurich, Institute for Particle Physics, Zurich, Switzerland}}
\newcommand{\INSTEG}{\affiliation{University of Geneva, Section de Physique, DPNC, Geneva, Switzerland}}
\newcommand{\INSTDG}{\affiliation{H. Niewodniczanski Institute of Nuclear Physics PAN, Cracow, Poland}}
\newcommand{\INSTCB}{\affiliation{High Energy Accelerator Research Organization (KEK), Tsukuba, Ibaraki, Japan}}
\newcommand{\INSTED}{\affiliation{Institut de Fisica d'Altes Energies (IFAE), Bellaterra (Barcelona), Spain}}
\newcommand{\INSTEC}{\affiliation{IFIC (CSIC \& University of Valencia), Valencia, Spain}}
\newcommand{\INSTEI}{\affiliation{Imperial College London, Department of Physics, London, United Kingdom}}
\newcommand{\INSTGF}{\affiliation{INFN Sezione di Bari and Universit\`a e Politecnico di Bari, Dipartimento Interuniversitario di Fisica, Bari, Italy}}
\newcommand{\INSTBE}{\affiliation{INFN Sezione di Napoli and Universit\`a di Napoli, Dipartimento di Fisica, Napoli, Italy}}
\newcommand{\INSTBF}{\affiliation{INFN Sezione di Padova and Universit\`a di Padova, Dipartimento di Fisica, Padova, Italy}}
\newcommand{\INSTBD}{\affiliation{INFN Sezione di Roma and Universit\`a di Roma ``La Sapienza'', Roma, Italy}}
\newcommand{\INSTEB}{\affiliation{Institute for Nuclear Research of the Russian Academy of Sciences, Moscow, Russia}}
\newcommand{\INSTHA}{\affiliation{Kavli Institute for the Physics and Mathematics of the Universe (WPI), Todai Institutes for Advanced Study, University of Tokyo, Kashiwa, Chiba, Japan}}
\newcommand{\INSTCC}{\affiliation{Kobe University, Kobe, Japan}}
\newcommand{\INSTCD}{\affiliation{Kyoto University, Department of Physics, Kyoto, Japan}}
\newcommand{\INSTEJ}{\affiliation{Lancaster University, Physics Department, Lancaster, United Kingdom}}
\newcommand{\INSTFC}{\affiliation{University of Liverpool, Department of Physics, Liverpool, United Kingdom}}
\newcommand{\INSTFI}{\affiliation{Louisiana State University, Department of Physics and Astronomy, Baton Rouge, Louisiana, U.S.A.}}
\newcommand{\INSTJ}{\affiliation{Universit\'e de Lyon, Universit\'e Claude Bernard Lyon 1, IPN Lyon (IN2P3), Villeurbanne, France}}
\newcommand{\INSTHB}{\affiliation{Michigan State University, Department of Physics and Astronomy,  East Lansing, Michigan, U.S.A.}}
\newcommand{\INSTCE}{\affiliation{Miyagi University of Education, Department of Physics, Sendai, Japan}}
\newcommand{\INSTDF}{\affiliation{National Centre for Nuclear Research, Warsaw, Poland}}
\newcommand{\INSTFJ}{\affiliation{State University of New York at Stony Brook, Department of Physics and Astronomy, Stony Brook, New York, U.S.A.}}
\newcommand{\INSTGJ}{\affiliation{Okayama University, Department of Physics, Okayama, Japan}}
\newcommand{\INSTCF}{\affiliation{Osaka City University, Department of Physics, Osaka, Japan}}
\newcommand{\INSTGG}{\affiliation{Oxford University, Department of Physics, Oxford, United Kingdom}}
\newcommand{\INSTBB}{\affiliation{UPMC, Universit\'e Paris Diderot, CNRS/IN2P3, Laboratoire de Physique Nucl\'eaire et de Hautes Energies (LPNHE), Paris, France}}
\newcommand{\INSTGC}{\affiliation{University of Pittsburgh, Department of Physics and Astronomy, Pittsburgh, Pennsylvania, U.S.A.}}
\newcommand{\INSTFA}{\affiliation{Queen Mary University of London, School of Physics and Astronomy, London, United Kingdom}}
\newcommand{\INSTE}{\affiliation{University of Regina, Department of Physics, Regina, Saskatchewan, Canada}}
\newcommand{\INSTGD}{\affiliation{University of Rochester, Department of Physics and Astronomy, Rochester, New York, U.S.A.}}
\newcommand{\INSTBC}{\affiliation{RWTH Aachen University, III. Physikalisches Institut, Aachen, Germany}}
\newcommand{\INSTFB}{\affiliation{University of Sheffield, Department of Physics and Astronomy, Sheffield, United Kingdom}}
\newcommand{\INSTDI}{\affiliation{University of Silesia, Institute of Physics, Katowice, Poland}}
\newcommand{\INSTEH}{\affiliation{STFC, Rutherford Appleton Laboratory, Harwell Oxford,  and  Daresbury Laboratory, Warrington, United Kingdom}}
\newcommand{\INSTCH}{\affiliation{University of Tokyo, Department of Physics, Tokyo, Japan}}
\newcommand{\INSTBJ}{\affiliation{University of Tokyo, Institute for Cosmic Ray Research, Kamioka Observatory, Kamioka, Japan}}
\newcommand{\INSTCG}{\affiliation{University of Tokyo, Institute for Cosmic Ray Research, Research Center for Cosmic Neutrinos, Kashiwa, Japan}}
\newcommand{\INSTGI}{\affiliation{Tokyo Metropolitan University, Department of Physics, Tokyo, Japan}}
\newcommand{\INSTF}{\affiliation{University of Toronto, Department of Physics, Toronto, Ontario, Canada}}
\newcommand{\INSTB}{\affiliation{TRIUMF, Vancouver, British Columbia, Canada}}
\newcommand{\INSTG}{\affiliation{University of Victoria, Department of Physics and Astronomy, Victoria, British Columbia, Canada}}
\newcommand{\INSTDJ}{\affiliation{University of Warsaw, Faculty of Physics, Warsaw, Poland}}
\newcommand{\INSTDH}{\affiliation{Warsaw University of Technology, Institute of Radioelectronics, Warsaw, Poland}}
\newcommand{\INSTFD}{\affiliation{University of Warwick, Department of Physics, Coventry, United Kingdom}}
\newcommand{\INSTGE}{\affiliation{University of Washington, Department of Physics, Seattle, Washington, U.S.A.}}
\newcommand{\INSTGH}{\affiliation{University of Winnipeg, Department of Physics, Winnipeg, Manitoba, Canada}}
\newcommand{\INSTEA}{\affiliation{Wroclaw University, Faculty of Physics and Astronomy, Wroclaw, Poland}}
\newcommand{\INSTH}{\affiliation{York University, Department of Physics and Astronomy, Toronto, Ontario, Canada}}
 
\INSTC
\INSTEE
\INSTFE
\INSTD
\INSTGA
\INSTI
\INSTGB
\INSTFG
\INSTFH
\INSTBA
\INSTEF
\INSTEG
\INSTDG
\INSTCB
\INSTED
\INSTEC
\INSTEI
\INSTGF
\INSTBE
\INSTBF
\INSTBD
\INSTEB
\INSTHA
\INSTCC
\INSTCD
\INSTEJ
\INSTFC
\INSTFI
\INSTJ
\INSTHB
\INSTCE
\INSTDF
\INSTFJ
\INSTGJ
\INSTCF
\INSTGG
\INSTBB
\INSTGC
\INSTFA
\INSTE
\INSTGD
\INSTBC
\INSTFB
\INSTDI
\INSTEH
\INSTCH
\INSTBJ
\INSTCG
\INSTGI
\INSTF
\INSTB
\INSTG
\INSTDJ
\INSTDH
\INSTFD
\INSTGE
\INSTGH
\INSTEA
\INSTH

\author{K.\,Abe}\INSTBJ
\author{J.\,Adam}\INSTFJ
\author{H.\,Aihara}\INSTCH\INSTHA
\author{T.\,Akiri}\INSTFH
\author{C.\,Andreopoulos}\INSTEH
\author{S.\,Aoki}\INSTCC
\author{A.\,Ariga}\INSTEE
\author{S.\,Assylbekov}\INSTFG
\author{D.\,Autiero}\INSTJ
\author{M.\,Barbi}\INSTE
% ...

\collaboration{The T2K Collaboration}\noaffiliation

\date{\today}

\begin{abstract}
The T2K experiment has performed a search for $\nue$ disappearance due to sterile neutrinos
using $5.9 \times 10^{20}$ protons on target 
for a baseline of $280\m$ in a 
neutrino beam peaked at about $500\mev$. A sample of \nue CC interactions in the off-axis near 
detector has been selected with a purity of 63\% and an efficiency of 26\%.  
The p-value for the null hypothesis is 0.085 and the excluded region at 95\% CL 
is approximately $\stee > 0.3$ for $\dmsqfo > 7 ~\ev^2 / \c^4$.
\newpage

\end{abstract}

\pacs{Valid PACS appear here}

\maketitle

 \textit{Introduction ---} In the last two decades, several experiments have observed neutrino 
oscillations compatible with the hypothesis of neutrino mixing in a three active flavours basis, 
described by the PMNS matrix~\cite{MNS}. Nevertheless, there exist experimental data that cannot be 
accommodated in this framework: the deficit of $\nue$ originating from intense radioactive sources 
in the calibration of the solar neutrino gallium detectors SAGE~\cite{SAGE1,SAGE2} and 
GALLEX~\cite{GALLEX} and $\nueb$ rates near nuclear reactors~\cite{RAA}. Those experiments cover 
$L/E$ values of order $1\m/\mev$, where $L$ is the neutrino flight-path and $E$ is the neutrino 
energy, too large to observe any sizeable effect for the standard neutrino mass differences.
These anomalies can be interpreted as neutrino oscillations if the PMNS matrix is extended by introducing
a new sterile neutrino $\nst$ (3+1 model) with a mass of order $1\ev/\c^2$ 
\cite{RAA,RAAGAL}. The deficit would be due to $\stackrel{_{(-)}}{\nue} \to \nu_s$ oscillations. The \nue beam 
component is studied at the ND280 near detectors of the T2K experiment~\cite{Abe:2011ks} to 
search for \nue disappearance. The analysis presented here considers $\nue\to\nst$ oscillations, 
given by the \nue survival probability in the approximation of two neutrino mass states:
\begin{linenomath}
\begin{align}
P(\stackrel{_{(-)}}{\nue} \-> \stackrel{_{(-)}}{\nue}) = 1 - \stee 
\sin^2\left(1.267\dfrac{{\dmsqfo} L}{E}\right)
\label{eq:psurv}
\end{align}
\end{linenomath}
where $\stee$ is the oscillation amplitude, ${\dmsqfo}[\ev^2 / \c^4] $ is the mass squared difference 
between the new sterile mass state and the weighted average of the active standard mass states, 
with $L[\m]$ and $E[\mev]$.

While anomalous excesses that might be explained by $\stackrel{_{(-)}}{\nue}$ appearance 
through sterile mixing have been observed by the MiniBooNE~\cite{PhysRevLett.105.181801} and 
LSND~\cite{Aguilar:2001ty} experiments, an explanation of all anomalies as sterile oscillations is disfavoured due 
to tension between appearance and disappearance data~\cite{GLOBALFIT_STERILES_1,GLOBALFIT_STERILES_2,GLOBALFIT_STERILES_3,GIUNTI}. In the absence of a consensus candidate model, new probes using the simple 3+1 model may be able to provide some insights into the existing anomalies. This analysis assumes no \num disappearance or \nue appearance.

With the given combination of $L$ and $E$,
this analysis is sensitive to \nue disappearance for $\dmsqfo\gtrsim2\ev^2  / \c^4$ in a sample of 
$\nue$ charged current (CC) interactions~\cite{NUE_PAPER}. A likelihood ratio fit to the  
reconstructed neutrino energy spectrum of the $\nue$ CC interactions is used to test the sterile 
neutrino hypothesis. A high purity sample of photon conversions from $\pi^0$ decays is included in 
the fit to control the dominant background in the $\nue$ sample. In addition, a selection of 
$\num$ CC interactions at ND280 is used to constrain the neutrino flux and cross section 
uncertainties in order to substantially reduce the uncertainties on the predicted 
$\nue$CC interaction rate.

 \textit{The T2K experiment ---} The T2K experiment uses a neutrino beam produced at the J-PARC 
facility in Japan to study neutrino oscillations and neutrino interactions 
~\cite{Abe:2011ks}. 
Electron and muon neutrinos are produced from the decay of pions 
and kaons generated when a $30\gev$ proton beam impinges on a graphite target. The detector ND280 
sits 280\m from the proton target $2.5^{\circ}$ from the primary proton beam direction (off-axis) 
and observes interactions of neutrinos from the beam, whose \nue component is peaked at an energy of 500\mev. 
The present analysis uses neutrino interactions on polystyrene scintillator or water inside two Fine Grained Detectors 
(FGDs~\cite{Amaudruz:2012pe}) that corresponds to a total fiducial mass of about $1.6t$. Three Time Projection Chambers (TPCs~\cite{Abgrall:2010hi}) adjacent 
to the FGDs are used to identify particle type and momentum. Electromagnetic calorimeters 
(ECal~\cite{Allan:2013ofa}) that surround the FGDs/TPCs (the Tracker) along the beam direction 
(Barrel ECal) and downstream (DsECal) additionally separate electron showers from muon tracks. The 
$\pi^0$ detector (P0D~\cite{Assylbekov201248}) is located upstream of the Tracker region and is 
used to veto interactions outside the FGDs in this analysis.

The results presented in this analysis are based on data taken from January 2010 to May 2013 which
corresponds to a total exposure at ND280 of $5.9\times10^{20}$ protons on target (POT) with a horn
configuration that enhances neutrinos and suppresses anti-neutrinos.

\begin{figure}
\begin{center}
\psfig{figure=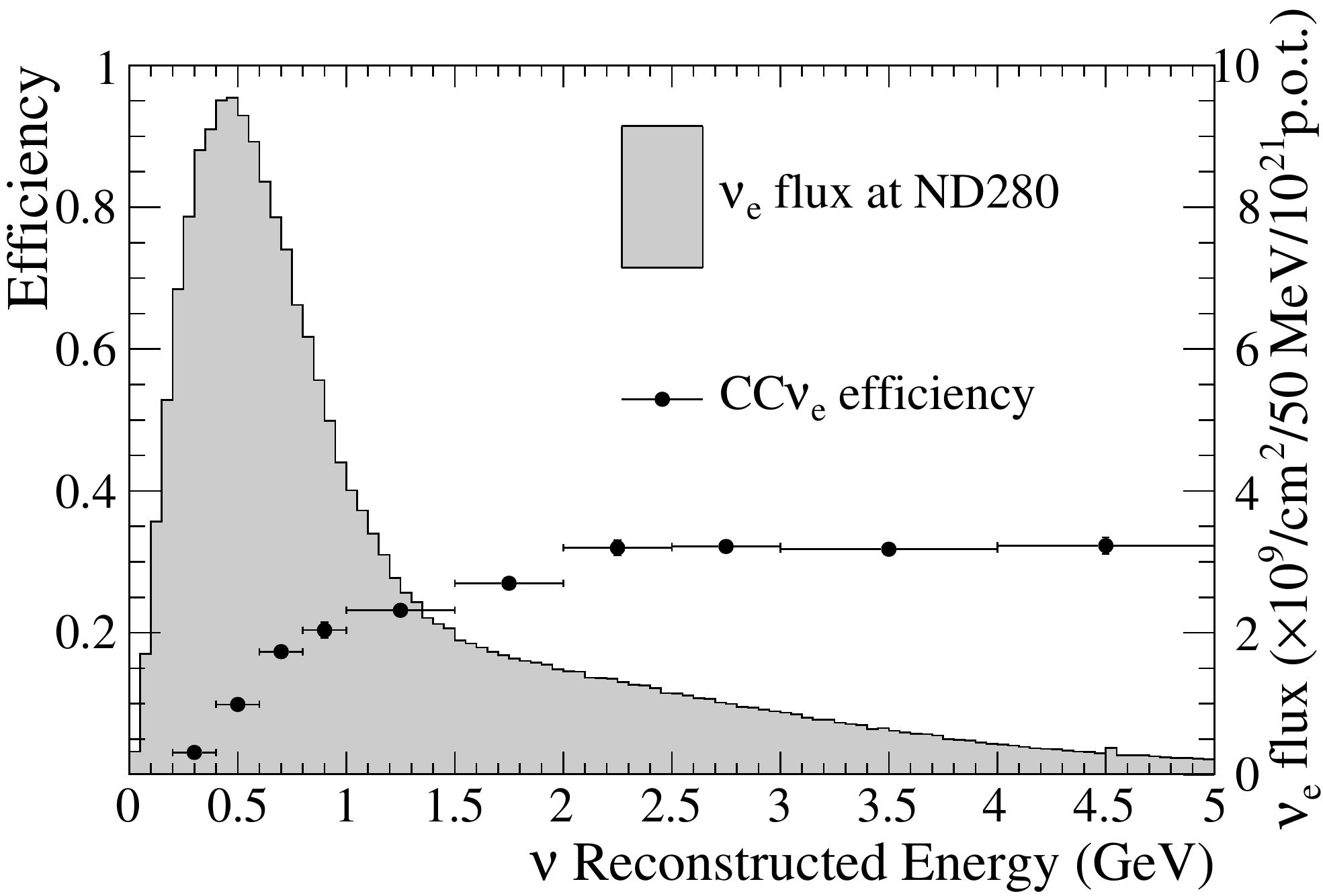,width=8cm,height=6cm}
\caption{Expected \nue flux at ND280 and CC \nue selection efficiency as a 
function of the true neutrino energy are shown. 
}
\label{fig:fluxpsurv}
\end{center}
\end{figure}

\begin{figure}
\begin{center}
\psfig{figure=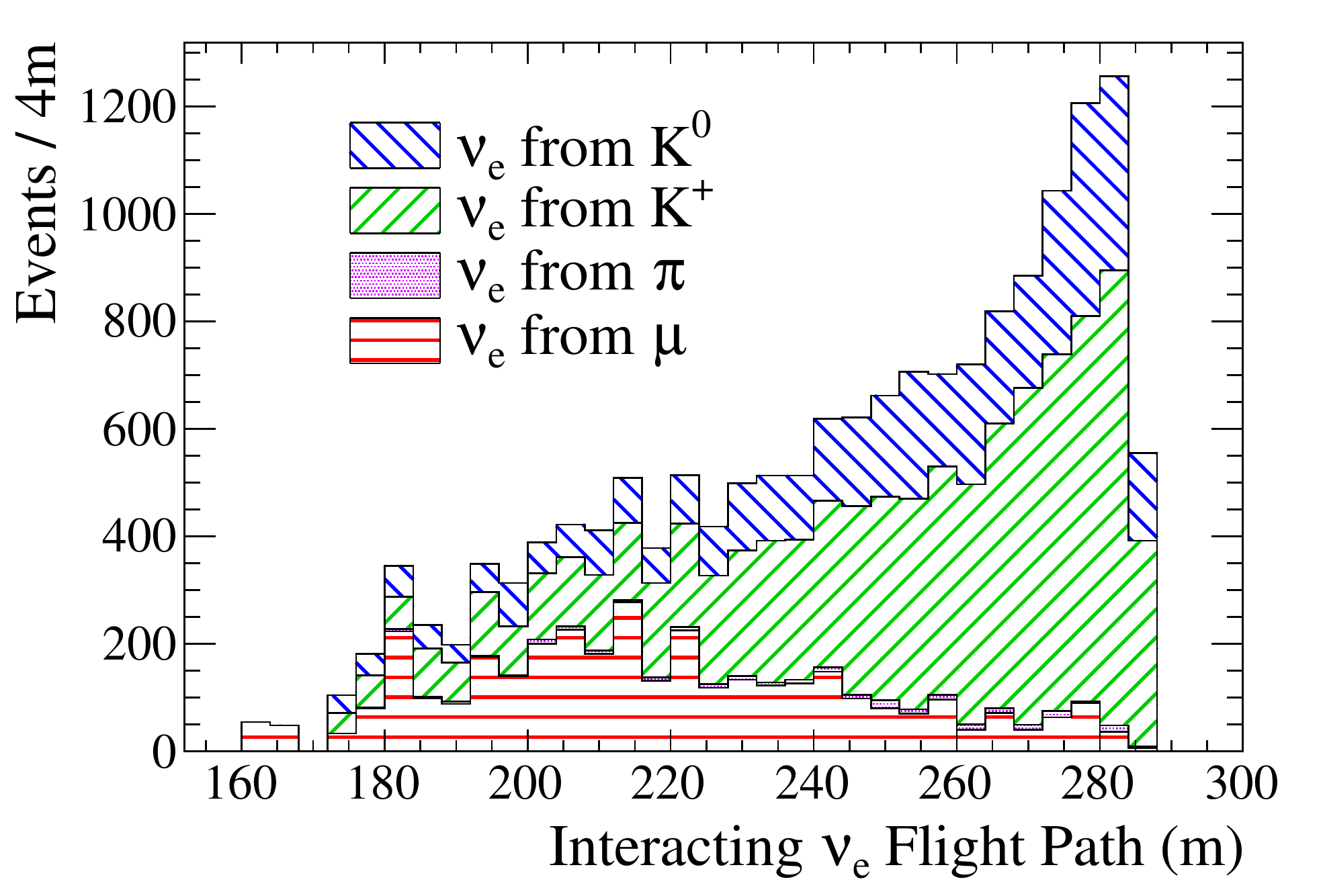,width=9cm}
\caption{
Expected neutrino flight path for \nue interacting in the ND280 FV, broken down by the neutrino
parent meson.}
\label{fig:fpath}
\end{center}
\end{figure}

\textit{$\nue$ flux at ND280 ---} The T2K beam is composed mostly of \num with 8.8\% \numb,
1.1\% \nue and $0.1\%$ \nueb ~\cite{T2K_FLUX}. The \nue flux at ND280 as a function of the neutrino 
energy is shown in \Cref{fig:fluxpsurv}. 
The fluxes of \nue and \nueb are produced predominantly by $K^{\pm}$ and $K^0$ decays at high energies ($E>1\gev$), and 
at low energies ($E<1\gev$) mainly by $\mu$ decay in flight~\cite{T2K_FLUX}.
$K^{\pm}$ and $K^0$ tend to decay near the hadron production 
point due to their short mean lifetime, while $\mu$ decay throughout the 96\m long decay volume, 
with a nearly flat decay length distribution. The \nue flight path distribution at ND280 is shown 
in \Cref{fig:fpath}. The average neutrino flight path, for \nue selected in the analysis, is 
$244\m$. The fluxes at the near detectors are predicted using a full Monte-Carlo simulation of the 
beam-line and modeling of hadron production cross section based on experimental data from 
NA61/SHINE~\cite{Abgrall:2011ae,Abgrall:2011ts}. The uncertainties on the \nue and \nueb fluxes 
range from 10\% to 20\% as a function of energy, prior to using any additional information from 
the \num CC interactions at ND280.

\textit{$\nu$ interactions at ND280 ---} The target material of the \nue CC selection in the Tracker 
is either water or polystyrene scintillator. At T2K energies, the dominant CC interaction is 
the CC quasi-elastic (CCQE) scattering off neutrons $(\nu_l n \to l^- p)$, where a negative lepton 
$l^-$ of the same flavour as the neutrino is created. 
At higher energies, neutrino CC interactions with pion production can take place
Those are CC resonant single $\pi$ production (CCRES),
coherent $\pi$ production (CCCoh) and multi-$\pi$ production due to deep inelastic scattering 
(CCDIS). As the \num flux is much larger than the \nue flux, the relative rate of $\num$ CC
interactions is expected to be $\sim100$ times larger than the analysis signal, $\nue$ CC 
interactions. Event selections in the Tracker are designed to enhance the selection of $\nu$-carbon or $\nu$-oxygen 
interactions inside the FGD fiducial volume (FV).

The most important background for \nue interactions are $\stackrel{_{(-)}}{\num}$ CCDIS or Neutral 
Current (NC) interactions which produce a $\pi^0$ ($\num N\to \pi^0 X$). The $\pi^0$ predominantly decays to two
photons and any electrons produced within the FV by $\gamma \to e^+e^-$ may be misidentified as
originating from $\nue$ CC interactions. Electrons in the FV may come from photons produced in $\nu$ interactions outside
the FV (OOFV) or inside it.

The neutrino event generator NEUT \cite{NEUT} simulates the neutrino interactions at ND280. 
Uncertainties in the neutrino-nucleus cross section models and re-interactions of pions within the 
nucleus (\textit{final state interaction}, FSI) are estimated by comparing the NEUT prediction 
with external neutrino, pion and electron scattering data~\cite{Abe:2013xua}. Each cross section is
characterized using a minimal set of parameters with large prior uncertainties between 20\% and 40\%.

\textit{Flux and cross section constraints at ND280 ---} Assuming no \num disappearance,
a measurement of \num CC interactions at ND280 is used to reduce the flux and the cross 
section uncertainties in the \nue signal prediction.
This is possible due to the significant correlation between the \num and \nue fluxes, originating 
from decays of the same hadron types. A similar technique is used in other T2K
measurements~\cite{Abe:2013hdq,PhysRevLett.112.092003}. Possible differences between \nue and \num 
cross sections of up to $3\%$, due to radiative corrections or differences in the nucleon form 
factors~\cite{Day:2012gb}, are included as a systematic error. 

A predominantly $\num$ CC 
interaction event sample is selected by identifying the highest momentum negative track originating within the 
FV which is compatible with a muon. This is done by exploiting the tracking and particle 
identification capabilities of the TPCs. Based on the presence of charged pions, the $\num$CC sample 
is further separated into three categories: events without pions (CC-0$\pi$), events with one $\pi^+$ 
(CC-$\pi^+$) and other interactions which produce a $\pi^-$, $\pi^0$ or more than one pion (CC-Oth). This 
provides sensitivity to the rate of \num CCQE, CCRES and CCDIS interactions. The three samples are 
binned in muon momentum and angle and they are fit to evaluate the neutrino flux and cross section uncertainties 
that are used as prior uncertainties in the \nue disappearance analysis.

\textit{Electron neutrino selection at ND280 and systematic uncertainties ---}
%
%
A sample of $\nu_e$ CC events  
is obtained by selecting electron-like events 
with the most energetic negatively charged track starting either in the FGD1 or  FGD2 FV. 
Electron candidates are selected by combining the particle identification (PID) capabilities of
the TPCs and ECals
to reject 99.8\% of muons. 
$\pi^0$
backgrounds are reduced by rejecting events where a positive electron-like track  is identified within 100 mm of the electron candidate 
and the $e^+e^-$ invariant mass is smaller than 100 $\mbox{MeV/c}^2$. 
Additionally we require that there are no tracks in the detectors upstream of the interaction vertex 
to reject $\nu_{\mu} N \to \pi^0 X$ interactions outside the FV.
%
%
%
$\nu_e$ CC interactions are selected with an overall efficiency of 26\% (see \Cref{fig:fluxpsurv}) and a purity of 63\%. 
The majority of the background (72\%) is electrons from 
conversion of $\pi^0$ decay photons 
($\nu_{\mu} N \to \pi^0 X$). 
The remaining background is from neutrino 
interactions where muons ($14\%$) or protons and pions ($14\%$)  
are misidentified as electrons. 
A significant component of the background (35\%) is due to particles produced outside the FV,
as in the magnet, dead materials of the FGDs and TPCs, ECal, P0D or surrounding material. 
Those neutrino interactions occur on heavier nuclei (e.g. iron, aluminium, lead)
with larger cross section uncertainties (30\%). 
This background is large at low energy.

%
%
A control sample is used to measure the $\nu_{\mu} N \to \pi^0 X$ background.
It is selected by requiring two electron-like tracks in the TPC with a common vertex in 
the FGD (distance between the starting points of the two tracks less than $10 ~\mbox{mm}$) and 
invariant mass less than $50 ~\mbox{MeV/c}^2$. 
%
%
The control sample has an overall selection efficiency with respect to the total number of photons converting in the FGDs of about 12\% 
and is a highly pure background sample predominantly consisting of 
photon conversion (92\%) from  $\nu_{\mu} N \to \pi^0 X$ in NC and CCDIS interactions. 
The kinematics of the photons in the control and signal samples are similar. 
Furthermore,  62\% of the control sample 
$\nu_{\mu}$ 
events are OOFV $\nu_{\mu} N \to \pi^0 X$, which provides a 
direct constraint for the $\nu_e$ sample background.
A more detailed description of the selection of both the $\nu_e$ and the  
control
samples is reported in \cite{NUE_PAPER}. 

\begin{figure}
\psfig{figure=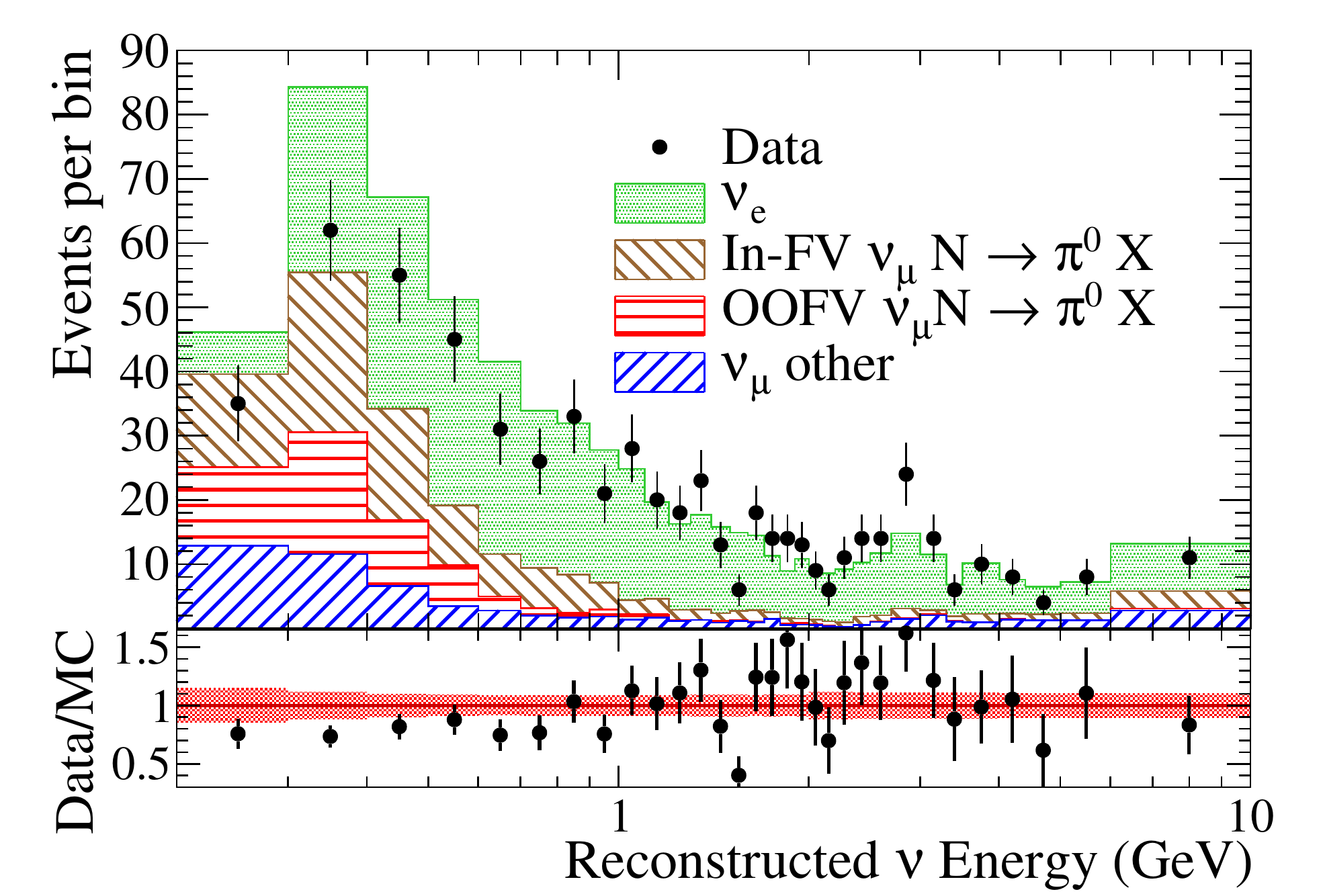,width=9cm}
\psfig{figure=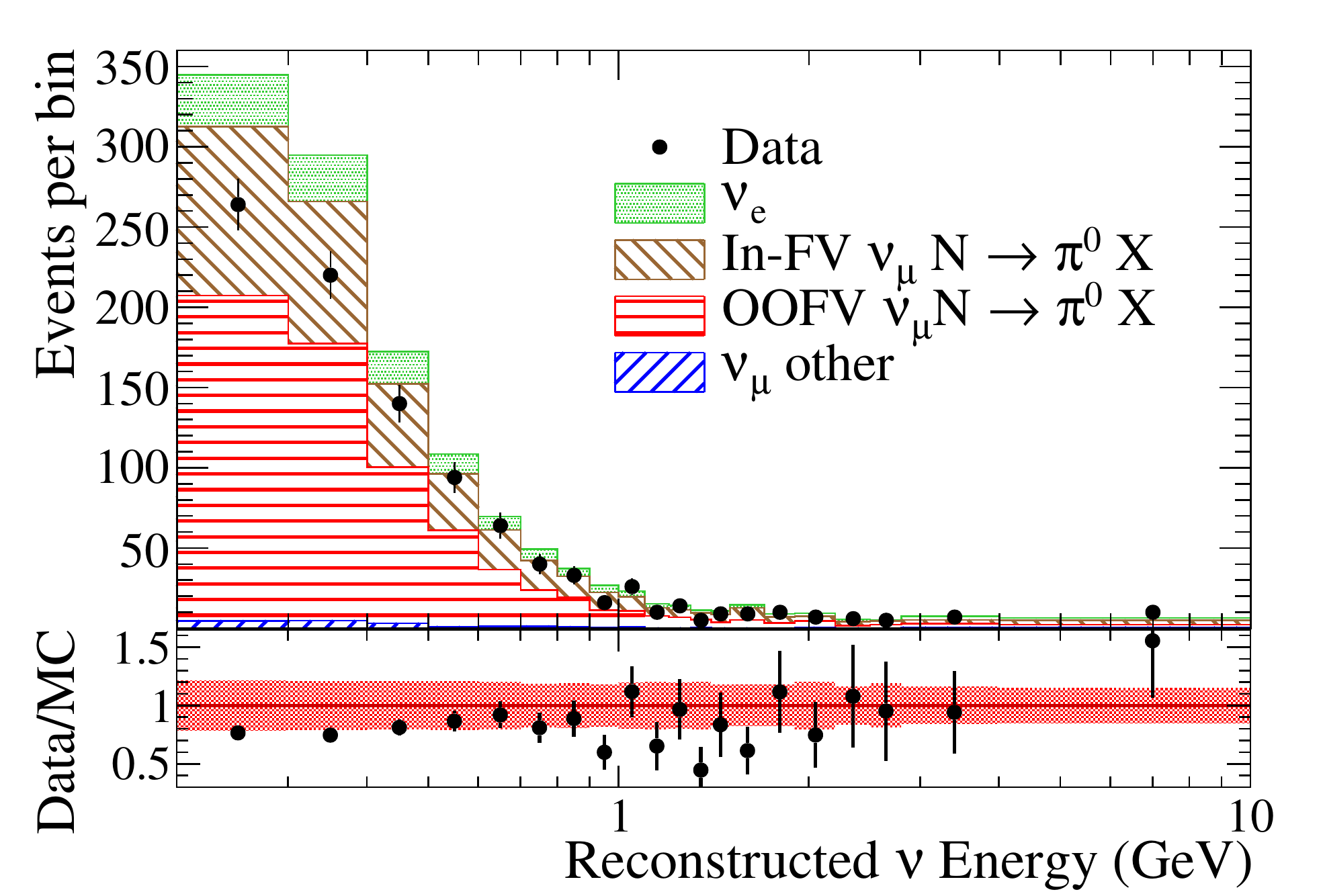,width=9cm}
\caption{\label{fig:ereco} Reconstructed energy distributions of the $\nu_e$ (top) and 
control
(bottom) samples.
The distributions are broken down by $\nu_e$ interactions (signal), background inside the fiducial volume due to $\nu_{\mu} N \to \pi^0 X$ (In-FV $\nu_{\mu} N \to \pi^0 X$), 
background outside the fiducial volume 
due to $\nu_{\mu} N \to \pi^0 X$ (OOFV $\nu_{\mu} N \to \pi^0 X$) and all other sources of background ($\nu_{\mu}$ other).
Both $\nu$ and $\bar{\nu}$ are included in the samples.
The ratio of the data to the MC expectation in the null oscillation hypothesis is shown for both samples. The red error band corresponds to the fractional systematic uncertainty.
Black dots represent the data with the statistical uncertainty.}
\end{figure}

The reconstructed $\nue$ energy spectrum ($E_{reco}$), assuming a CCQE interaction, is inferred from the outgoing electron candidate momentum and angle, as in \cite{Ereco}.
\nue disappearance would affect the rate and energy spectrum of $\nue$ CC interactions.
%
%
\Cref{fig:ereco} shows the $E_{reco}$ distributions of the $\nue$ and the 
control
samples. 
 %
%
A total of 614 $\nu_e$ CC candidates are selected in the $\nu_e$ sample and 
$665 \pm 51 ~(\mbox{syst})$ events are expected, assuming no oscillation 
and with the systematic uncertainties described below.
%
%
The number of selected events in the 
control
sample is 989 in data, with an expectation of $1236 \pm 246 ~(\mbox{syst})$.
%
%
Systematic uncertainties on the flux, cross section and detector response are taken into 
account using the approach adopted in \cite{NUE_PAPER}. 
The systematic uncertainties on the flux and $\nu_e$-$\nu_{\mu}$ common cross sections are 
constrained by fitting the $\nu_{\mu}$ CC sample as described earlier.
The unconstrained cross-section systematic uncertainties include several contributions:
the difference between the interaction cross section of $\nu_{\mu}$ and $\nu_e$, 
between $\nu$ and $\bar{\nu}$ and the uncertainty on OOFV interactions. 
FSI uncertainties contribute 1.5\% (2.7\%) to the $\nu_e$  ($\nu_{\mu} N \to \pi^0 X$) sample systematic uncertainty.
The detector systematic uncertainties have been evaluated independently for the TPCs, FGDs and ECal. 
The largest sources of uncertainties are given by the TPC momentum resolution and the PID. 
In Table \ref{tab:neventsyst}, 
the effect of each group of systematic uncertainties on the total expected number of signal and 
signal plus background events is shown.
In \Cref{fig:ereco} the effect of the systematic uncertainties on the $E_{reco}$ distributions is shown.  
The simulation overestimates the data in   
both the $\nu_e$ and 
control
sample distributions at low energy.
However this overestimation in the
control
sample is within one standard deviation of expectation.
\begin{table}
\caption{\label{tab:neventsyst} Fractional variation (RMS/mean in \%) of the expected total number of events for $\nu_e$ (all events and signal only) and 
control sample 
in the null oscillation hypothesis due to the effect of the systematic uncertainties. Existing correlations between systematics are taken into account.}
\begin{ruledtabular}
 \begin{tabular}{lccc}
 \multirow{2}{*}{Error source (\# param.)}  & \multicolumn{1}{c}{$\nu_e$ sample}  & \multicolumn{1}{c}{$\nu_e$ sample} & \multicolumn{1}{c}{control } \\
 						& \multicolumn{1}{c}{(sig+bkg)} & \multicolumn{1}{c}{(sig only)} & \multicolumn{1}{c}{sample} \\
\hline 
$\nu_{\mu}$ - $\nu_{e}$ common (40)           	& 4.4    & 5.2	& 6.7 \\
Unconstrained (5)           		& 3.7    & 3.0	& 17.8 \\
Detector + FSI (10)           		& 5.1    & 5.5	& 5.5 \\
\hline
Total (55)					& 7.6    & 8.1 	& 19.9 \\
\end{tabular}
\end{ruledtabular}
\end{table}

\textit{Oscillation fit ---}
The sterile oscillation parameters $\stee$ and $\dmsqfo$ are estimated with a Poisson binned likelihood ratio method.
The expected reconstructed neutrino energy distributions are compared to data with a simultaneous fit to the selected $\nu_e$ and 
control samples.
The range of $E_{reco}$ is from $\unit[0.2]{GeV}$ to $\unit[10]{GeV}$.
The oscillation amplitude $\stee$ is restricted to the physical region.
The effect of systematic uncertainties is included in the fit with nuisance parameters (55 in total) constrained by a Gaussian penalty term.
The oscillation probability \Cref{eq:psurv} affects $\nu_e$ signal events based on the true 
neutrino energy and flight path.

\begin{figure}
\psfig{figure=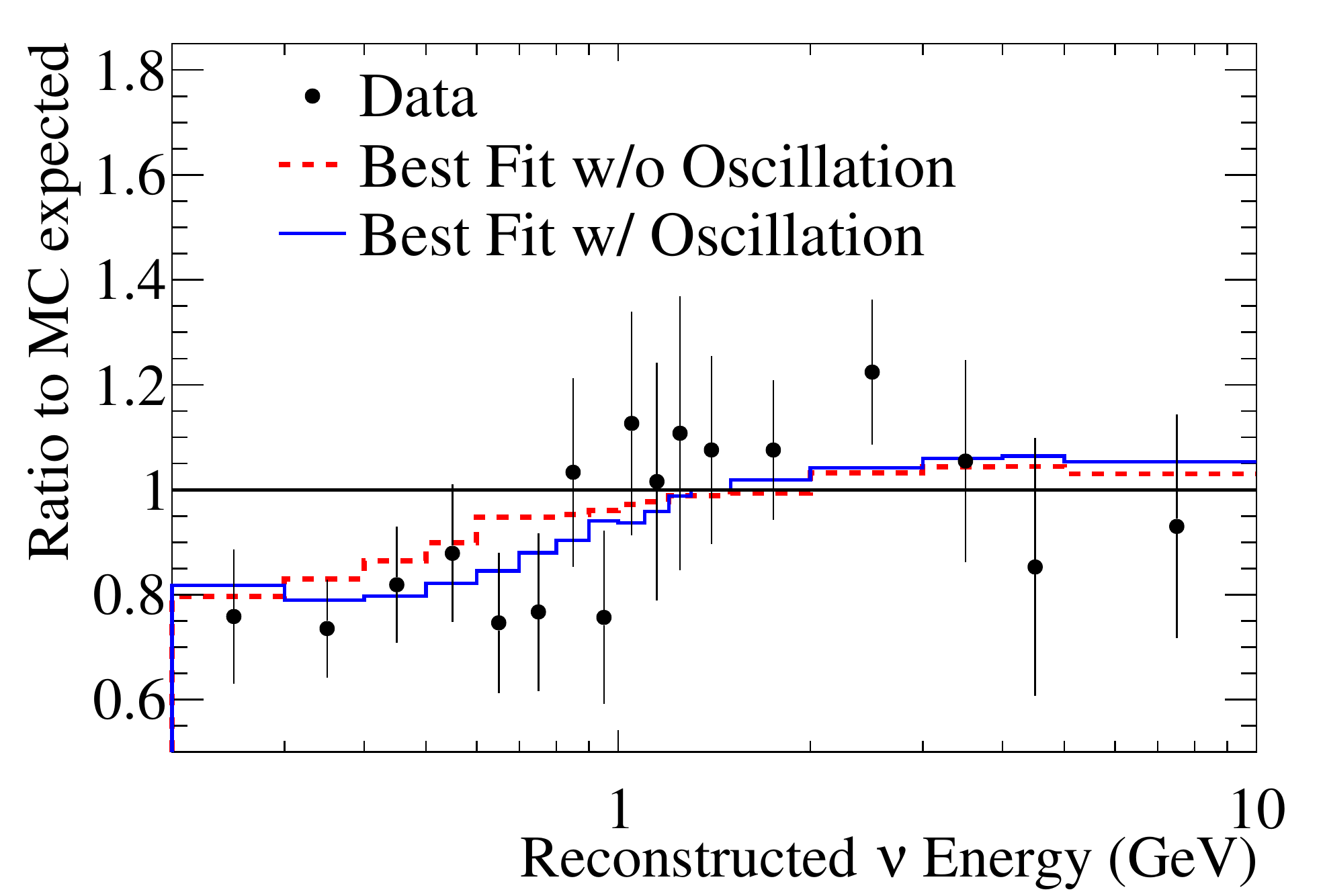,width=8cm}
\psfig{figure=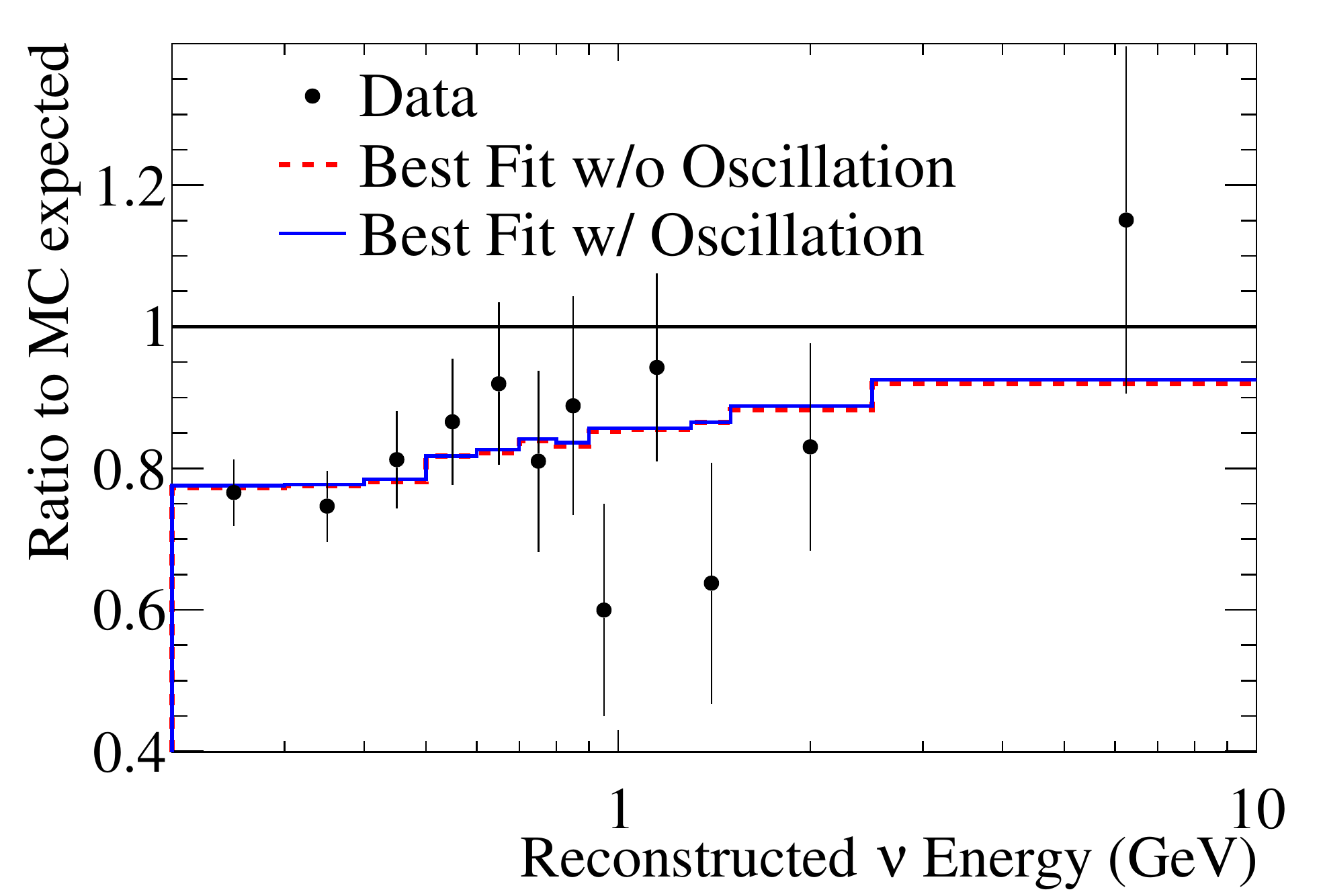,width=8cm}
\caption{\label{fig:ratio} 
The ratio of the best fit spectrum to the expected MC distribution, where the fit includes nuisance and oscillation parameters (blue) and nuisance parameters only (red dashed), is shown. 
The plots show the $\nu_e$ sample (top) and the 
control
sample (bottom).
The black line corresponds to the expected non-oscillated MC before the fit.
The black dots show the data. Statistical uncertainties are shown. 
}
\end{figure}

The best-fit oscillation parameters are $\stee = 1$ and $\dmsqfo = 2.05 ~\ev^2 / \c^4$. 
The $\chi^2 / ndf$ is 42.16/49.
Most of the best-fit systematic parameters are within a $0.5\sigma$ deviations and always within $1\sigma$ from the prior values.
The systematic parameter corresponding to the normalization of the $\nu_{\mu} N \to \pi^0 X$ OOFV component is reduced by 31\% 
($\sim1\sigma$) due to the deficit at low energy in the 
control 
sample.
The ratio between the best-fit and the expected non-oscillated MC distributions is shown as a 
function of $E_{reco}$ for both the $\nu_e$ and the 
control
samples in \Cref{fig:ratio}. 
The best-fit, where the nuisance parameters
are allowed to float while the oscillation parameters are fixed to null, is also shown.
The corresponding $\chi^2 / ndf$ is 45.86/51.

\begin{figure}
\psfig{figure=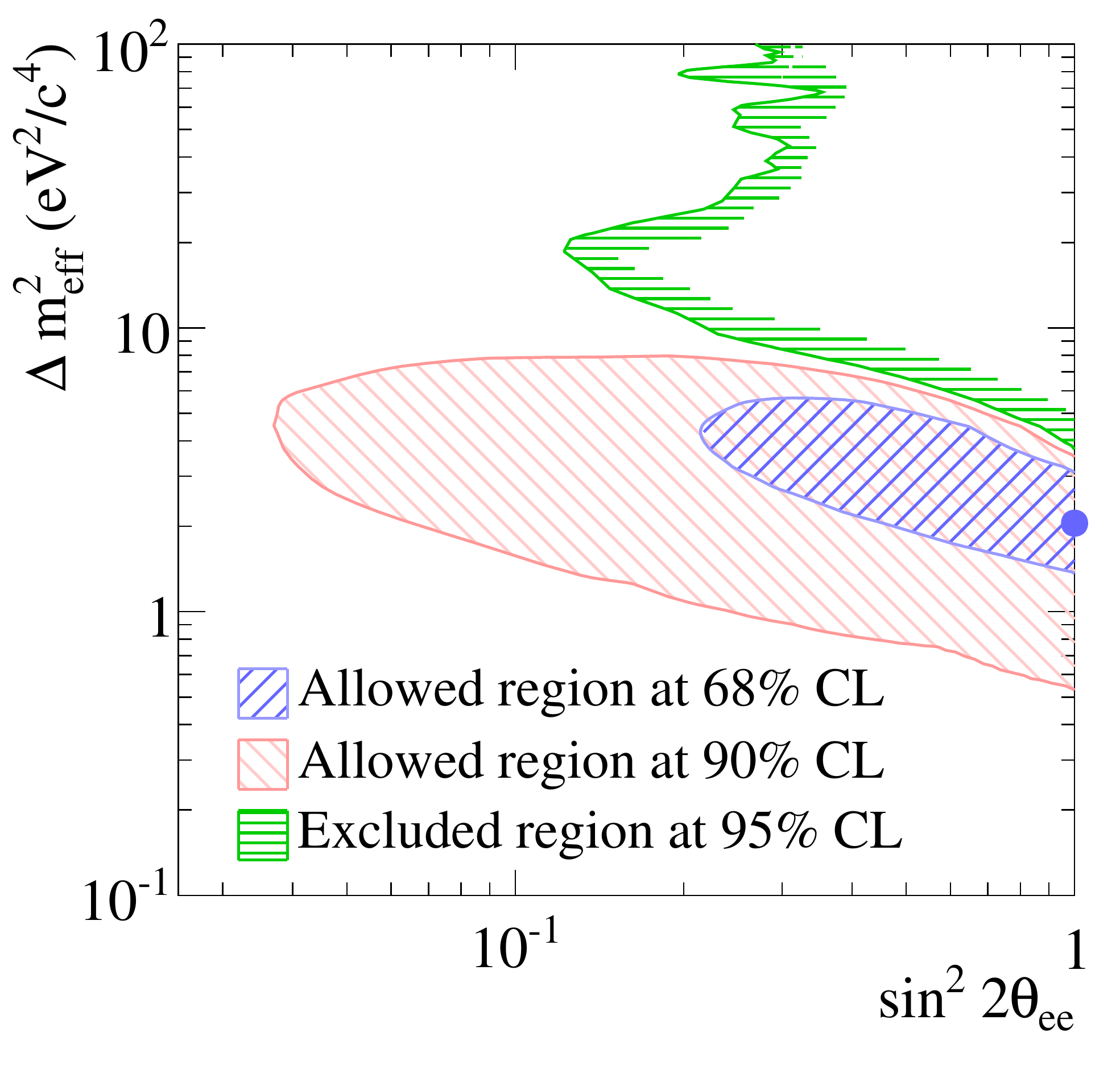,width=8cm}
\caption{\label{fig:exclusion} 68\% and 90\% CL allowed regions and 95\% CL exclusion region for the 
$ \stee \mbox{ - } \dmsqfo$ parameters measured with the T2K near detector. }
\end{figure}

The two-dimensional confidence intervals in the $\stee \mbox{ - } \dmsqfo$ parameter space are 
computed using the Feldman-Cousins method \cite{PhysRevD.57.3873}. 
The systematic uncertainties are incorporated using the method described in \cite{cousins-highland}.
The 68\%, 90\% and 95\% confidence regions are shown in \Cref{fig:exclusion}. 
The exclusion region at 95\% CL is approximately given by $\stee > 0.3 $ and $\dmsqfo > 7 ~\ev^2 / \c^4 $.

The $p$-value of the null oscillation hypothesis, computed using a profile likelihood ratio as a test statistic, is 0.085.

The impact of \num disappearance and \nue appearance on the present result is estimated by 
considering a non-null $\sin^2 2 \theta_{\mu \mu}$ in the 3+1 model. 
For $\sin^2 2 \theta_{\mu \mu}$ between 0 and 0.05, approximately the region not excluded by other experiments
\cite{GLOBALFIT_STERILES_1,MB_Disapp},
the 95\%CL exclusion on $\sin^2 2 \theta_{ee}$ moves by less than 0.1

\begin{figure}
\psfig{figure=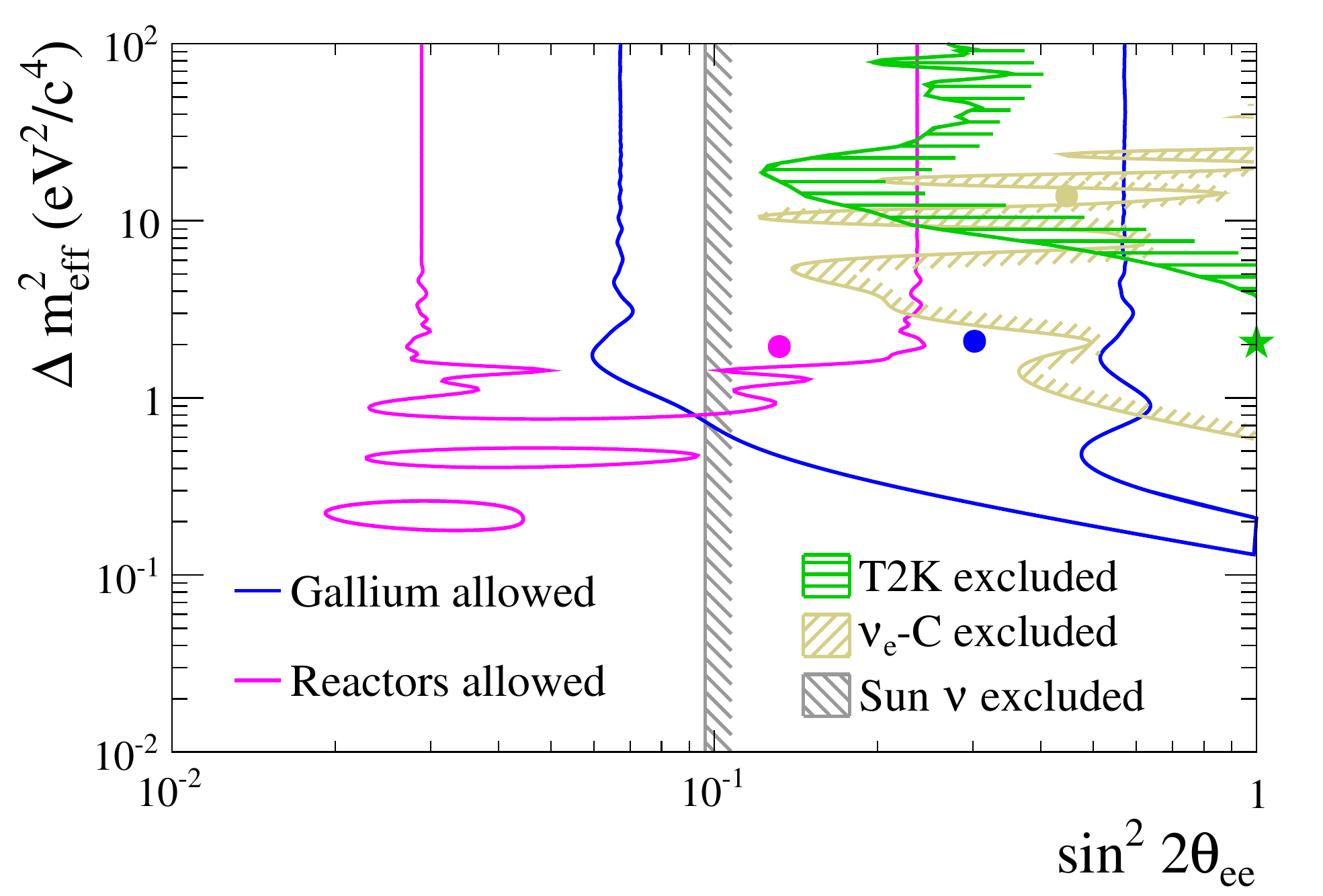,width=9cm,left}
\caption{\label{fig:giunti} The T2K excluded region in the $\stee $ - $ \dmsqfo$ parameter space 
at 95\% CL is compared with the other experimental results available in literature:
allowed regions of gallium and reactor anomalies and excluded regions by $\nu_e$-carbon interaction data and solar neutrino data \cite{GIUNTI}. 
The T2K best fit is marked by a green star; the best fit of other experimental results corresponds to circles of the same coloring as the limits.
}
\end{figure}

In  \Cref{fig:giunti} the T2K excluded region at 95\% CL is compared with $\nu_e$ disappearance 
allowed regions from the gallium anomaly and reactor anomaly.
The excluded regions from 
$\nu_e + ^{12}C \rightarrow ~^{12}N + e^-$ scattering
data of KARMEN \cite{nueC-KARMEN-1, nueC-KARMEN-2} and LSND \cite{nueC-LSND} experiments
and solar neutrino and KamLAND data \cite{solar-gallex, solar-Homestake, solar-SAGE, solar-SK, solar-SK-2, solar-SK-3, 
solar-SK-4, solar-SNO, solar-SNO-2, solar-SNO-3, solar-Borexino, solar-Borexino-2, solar-KAMLAND} 
are also shown.
The T2K result excludes part of the gallium anomaly and a small part of the reactor anomaly allowed regions.
The current T2K limit at 95\% CL is contained within the region excluded by the combined fit 
of the solar  and KamLAND data.

\vspace{1mm}
\textit{Conclusions ---}
T2K has performed a search for \nue disappearance with the near detector.
 The excluded region at 95\% CL is approximately $\stee > 0.3$ and $\dmsqfo > 7 ~\ev^2 / \c^4 $.
The p-value of the null oscillation hypothesis is 0.085.
Further data from T2K will reduce the statistical uncertainty, which is still an important limitation for the analysis.
%

\vspace{4mm}

We thank the J-PARC staff for superb accelerator performance and the
CERN NA61 collaboration for providing valuable particle production data.
We acknowledge the support of MEXT, Japan;
NSERC, NRC and CFI, Canada;
CEA and CNRS/IN2P3, France;
DFG, Germany;
INFN, Italy;
National Science Centre (NCN), Poland;
RSF, RFBR and MES, Russia;
MINECO and ERDF funds, Spain;
SNSF and SER, Switzerland;
STFC, UK; and
DOE, USA.
We also thank CERN for the UA1/NOMAD magnet,
DESY for the HERA-B magnet mover system,
NII for SINET4,
the WestGrid and SciNet consortia in Compute Canada,
GridPP, UK.
In addition participation of individual researchers
and institutions has been further supported by funds from: ERC (FP7), EU;
JSPS, Japan;
Royal Society, UK;
DOE Early Career program, USA.


\bibliographystyle{unsrt}


\end{document}